\documentclass[10pt,conference]{IEEEtran}
\usepackage{graphicx}
\usepackage{booktabs}
\usepackage{multirow}
\usepackage{amsmath}
\usepackage{amssymb}
\usepackage{siunitx}
\usepackage{newtxtext}
\usepackage[caption=false,font=footnotesize]{subfig}
\usepackage{enumitem}
\usepackage{microtype}
\usepackage{url}
\usepackage{xspace}
\usepackage[table]{xcolor}
\usepackage{tcolorbox}
\usepackage{array}
\usepackage{tabularx}
\usepackage{tikz}
\usetikzlibrary{arrows.meta,positioning,calc,fit,backgrounds}
\usepackage[ruled,lined,noend]{algorithm2e}
\SetAlgoNlRelativeSize{-1}
\SetAlgoSkip{smallskip}
\setlist[itemize]{leftmargin=*,noitemsep,topsep=2pt,parsep=0pt,partopsep=0pt}
\setlist[enumerate]{leftmargin=*,noitemsep,topsep=2pt,parsep=0pt,partopsep=0pt}
\SetAlFnt{\footnotesize}
\SetAlgoCaptionSeparator{.}

\usepackage[backend=biber,style=ieee,sorting=none,
            maxbibnames=4,minbibnames=1]{biblatex}
\AtBeginBibliography{\footnotesize}
\usepackage[hidelinks]{hyperref}

\usepackage{enumitem} % for \newlist and \setlist
\usepackage{amssymb}  % for \blacktriangleright

\newlist{rqlist}{itemize}{1}
\setlist[rqlist]{label=\small$\blacktriangleright$, leftmargin=1.5em, itemsep=6pt, parsep=4pt}

\newcommand{\dataset}{\textsc{DynFault}\xspace}
\newcommand{\pid}{\texttt{so\_id}\xspace}
\newcommand{\CF}{\textsc{CF}\xspace}  
\newcommand{\OF}{\textsc{OF}\xspace}  
\newcommand{\BA}{accuracy\xspace}
\newcommand{\ROCAUC}{ROC-AUC\xspace}
\newcommand{\PRAUC}{PR-AUC\xspace}
\newcommand{\tvmm}{\emph{T/V-mismatch}\xspace}
\newcommand{\kb}[1]{$k{=}#1$}
\newcommand{\ci}[2]{{\fontsize{6}{7}\selectfont[#1,\,#2]}}

\definecolor{tabhead}{HTML}{EDF1F6}   % light blue-gray header band
\definecolor{rowhi}{HTML}{E4EAF2}     % primary-row / highlight band
\definecolor{accent}{HTML}{1F3A5F}    % deep navy accent (grayscale-safe)
\newcommand{\headrow}{\rowcolor{tabhead}}   % shaded header band
\newcommand{\findrow}{\rowcolor{rowhi}}      % highlighted primary row
\tcbuselibrary{skins}
\newtcolorbox{rqbox}{
  enhanced,
  colback=gray!5,
  colframe=gray!30,
  boxrule=0.3pt,
  arc=1.5pt,
  leftrule=2.6pt,
  borderline west={2.6pt}{0pt}{accent},
  left=7pt,right=6pt,top=4pt,bottom=4pt,
  fontupper=\small
}

\begin{document}
\title{Evaluation-Strategy Gap in Fault Diagnosis of Deep Learning Programs}
\author{
    \IEEEauthorblockN{Sigma Jahan}
    \IEEEauthorblockA{Faculty of Computer Science \\
    Dalhousie University, Halifax, Canada \\
    sigma.jahan@dal.ca}
}

\maketitle

% -----------------------------------------------------------------------
\begin{abstract}
Deep Learning (DL) programs can fail during training for many reasons, and diagnosing the cause is a costly and time-consuming maintenance task. Techniques for diagnosing such failures are commonly assessed using within-program cross-validation, which may be inadequate for deployment settings involving previously unseen programs. It is therefore necessary to assess how performance differs across these settings and to identify the causes of any performance gap in established fault diagnosis techniques for DL. We investigate this gap using \dataset, a corpus of 5,542 fault-injected training traces from 38 real-world DL programs. We found a gap of 0.190 in balanced accuracy for existing fault diagnosis techniques between within-program evaluation and holding out whole programs. We also found the gap comes from program-level structure in the features, which led us to examine two runtime feature sets, curvature features and optimizer features, and their behavior on unseen programs. We found that curvature features are useful for instability detection on unseen programs, while optimizer and activation features help only on programs seen during training.
\end{abstract}


% -----------------------------------------------------------------------
\section{Introduction}
\label{sec:intro}

Deep Learning (DL) programs are prone to training-time failures caused by
configuration errors, implementation bugs, and numerical
instability~\cite{islam2019bugs,humbatova2020taxonomy}.
Diagnosing these failures is challenging, as DL programs can vary widely
in architecture, data set, and training configuration. The loss
landscape, the dynamics of the gradients and the convergence behavior of one program
can differ substantially from another~\cite{li2018visualizing}.
When a training run diverges, overfits, or shows inconsistent training and validation behavior, practitioners must diagnose the root cause from available information (e.g., runtime metrics, code snippets, model outputs), often under tight time and compute budgets~\cite{zhang2020empiricaldljobs,islam2020repairpatterns}. Fault diagnosis techniques based on dynamic analysis learn to identify such failures by observing runtime signals (e.g., loss trajectories, gradient statistics, and optimizer states)~\cite{wardat2022deepdiagnosis,cao2022deepfd,jahan2025default,qi2024coverage}.

Existing fault diagnosis techniques often report strong results under cross-validation within the program, and several also validate on a limited number of real-world fault benchmarks. For these techniques, each program
produces characteristic patterns in runtime metrics, so a classifier
trained in one set of programs can learn to identify those programs instead
of faults~\cite{geirhos2020shortcut}. Within-program splits put the same
programs in training and testing, and fixed benchmarks report one aggregate
score. Neither approach measures the impact on the performance of fault diagnosis for entirely new DL programs. Systematic evaluation on excluded groups is common practice in cross-project defect prediction~\cite{zimmermann2009cpdp,tantithamthavorn2017validation} as well as grouped cross-validation for clustered data~\cite{roberts_ecography2017,lyu_tosem2021}. However, similar evaluation strategies have not been widely adopted for fault diagnosis techniques that learn from runtime data, leading to two major gaps, as follows. 

\emph{\textbf{The impact of evaluation strategy on fault diagnosis remains unclear.}} We have limited systematic evidence on how much the performance of existing fault diagnosis techniques changes under different evaluation strategies, since previous work often overlooks these evaluation setups. A decline in performance could indicate that the fault signal does not transfer or that within-program evaluation partly rewards a classifier for recognizing the program rather than the fault~\cite{qi2024coverage}. These explanations imply different causes for the observed decline, which cannot be identified from the performance drop alone. Thus, more comprehensive evaluation and analysis are warranted to understand how the performance of fault diagnosis techniques reflects the overall diagnostic capability.

\emph{\textbf{The value of runtime features remains unclear across evaluation strategies.}} Existing fault diagnosis techniques differ in terms of the runtime metrics collected to learn fault patterns~\cite{cao2022deepfd, jahan2025default, autotrainer, wardat2022deepdiagnosis,schoop2021umlaut}. Richer instrumentation is often necessary to collect these metrics, which can be costly. However, it remains unclear whether the choice of evaluation strategy affects the usefulness of these runtime metrics. Thus, it is difficult to weigh the cost of richer logging against its likely benefit for unseen programs.

To address these gaps, we study \dataset, a corpus of 5,542 fault-injected training
traces from 38 real-world DL programs to compare two ways of splitting the
data. In \emph{within-program} evaluation, runs are split without regard to
which program they came from. In \emph{program-held-out} evaluation, all
runs from a given program stay together, so a tested program never appears
in training. We examine three diagnostic tasks, (a) fault-type
classification, (b) catastrophic-instability triage, and (c) training/validation
mismatch detection. In each task, we also compare two runtime feature
configurations, first-order optimizer and activation features, and second-order curvature features built from Hessian-vector products (HVPs). In short, we ask the following research questions.

\begin{rqlist}
  \item \textbf{RQ$\mathbf{_1}$: How large is the within-program vs.\ program-held-out gap for fault-type diagnosis, and what causes it?}
  We measure the gap on the four-class fault-type task and then run four controls (program-identity prediction, label permutation, per-program normalization, and distribution-level shift analysis). These controls trace the gap to program-level structure in the features. 
  \item \textbf{RQ$\mathbf{_2}$: When does catastrophic instability occur, and do curvature features detect it on unseen programs?}
  We find that instability occurs at initialization, before substantial training has taken place, and that curvature features improve detection on new programs at every observation window. They provide an instability signal that transfers.
  \item \textbf{RQ$\mathbf{_3}$: Do optimizer features generalize for training/validation mismatch detection?}
  We find that optimizer and activation features improve mismatch detection within a program but lose much of that benefit on new programs. The extra logging adds accuracy that does not transfer, and within-program evaluation hides the loss.
\end{rqlist}

We release \dataset and the full evaluation framework as a replication package~\cite{icsmeReplicationPackage}. Table~\ref{tab:qualitative_comparison} positions our study within existing fault diagnosis research and shows that program-held-out evaluation is rarely reported for learning-based techniques.

% --- Table: Qualitative comparison (moved from Related Work) ---
\begin{table}[htbp]
  \centering
  \footnotesize
  \caption{Fault diagnosis techniques for DL programs and their evaluation scope}
  \label{tab:qualitative_comparison}
  \setlength{\tabcolsep}{2.5pt}
  \resizebox{\columnwidth}{!}{%
  \begin{tabular}{@{}llclcc@{}}
    \toprule
    \headrow
    \textbf{Technique}
      & \textbf{Approach}
      & \textbf{Categories}
      & \textbf{Signals}
      & \textbf{Prog-held-out}
      & \textbf{Split} \\
    \midrule
    DeepFD~\cite{cao2022deepfd}
      & kNN/DT/RF & 5 types & Runtime & $\times$ & Within \\
    Qi et al.~\cite{qi2024coverage}
      & RF & 5 types & Runtime, coverage & $\times$ & Within \\
    AutoTrainer~\cite{autotrainer}
      & Rules & 5 problems & Runtime & --- & Within \\
    DeepDiagnosis~\cite{wardat2022deepdiagnosis}
      & Rules + DT & 8 symptoms & Runtime & --- & Within \\
    DEFault~\cite{jahan2025default}
      & Hierarchical RF & 7 types & Runtime, static & $\times$ & Within \\
    \midrule
    \findrow
    Our study
      & LR & 4--6 types & Runtime (\CF, \OF) & $\checkmark$ & Both \\
    \bottomrule
  \end{tabular}%
  }\\[2pt]
  \parbox{\columnwidth}{\scriptsize\raggedright
  $\checkmark$ evaluated under a program-held-out split, $\times$ not reported, --- not applicable (rule-based detectors).
  \emph{Categories} are the distinct fault types, training problems, or symptoms a technique reports.
  \emph{Signals}: runtime = per-epoch training metrics (e.g., loss, accuracy, gradient, weight). Coverage = neuron-coverage metrics. Static = code-structure features. \CF\ \& \OF\ = curvature and optimizer feature sets. Parenthesized numbers are reported feature counts.
  \emph{Approach}: kNN, DT = decision tree, RF = random forest, LR = logistic regression.}
\end{table}

% -----------------------------------------------------------------------

\section{Background and Corpus}
\label{sec:background}

\subsection{Dynamic Analysis Techniques for Fault Diagnosis}
\label{sec:tracediag}

Dynamic analysis techniques learn from per-epoch scalar metrics of a
training run, such as loss, accuracy, and gradient statistics, to
classify fault types or recommend corrective
actions~\cite{wardat2022deepdiagnosis,cao2022deepfd,jahan2025default,
qi2024coverage}.
Existing work evaluates under within-program splits
and, in several cases, on separate real-world fault benchmarks, but does not isolate how performance changes on a previously unseen program.
Our study addresses this dimension between programs by separately
characterizing two maintenance scenarios. Within-program reuse applies a
model to new runs of a program already in the training data, and
cross-program deployment applies it to a program not seen during training.

\subsection{The \dataset Corpus}
\label{sec:corpus}

\dataset is a corpus of 5,542 labeled training
traces constructed for this study.
Each trace is a per-epoch CSV of scalar training metrics produced by a
mutation-injected run of one of 38 distinct real-world DL programs
(FFNN, RNN, and CNN architectures). Each program is identified by its Stack Overflow post identifier (\pid).

\textbf{Baseline of the program and the mutation protocol.}
38 DL programs are real-world examples curated from Stack Overflow~\cite{jahan2025default}. We inject faults using the extended set of mutation operators, covering seven major fault
categories based on empirical studies of real DL
bugs~\cite{islam2019bugs,jahan2025default,yang2022dlframeworkbugs,deepcrime}.
These categories are hyperparameter, loss function, activation, layer
configuration, optimizer, weight initialization, and regularization.
We classify each mutant as killed or survived using the Generalized
Linear Model (GLM) criterion~\cite{nelder1972generalized} from previous work~\cite{jahan2025default, deepcrime}.
Mutants that are killed are assigned a fault label that reflects their mutation category. The Weight-initialization mutants did not produce killed runs, thus
six of the seven categories appear in our labeled corpus.

\textbf{Runtime feature configurations.}
Prior diagnosis techniques use standard metrics such as loss, accuracy, and gradient statistics~\cite{wardat2022deepdiagnosis,cao2022deepfd,jahan2025default}. \dataset also includes curvature metrics derived from Hessian-vector products (HVPs)~\cite{ghorbani2019investigation}, motivated by evidence that Hessian-derived metrics can localize training instability and fault sources in DL programs~\cite{jahan2025hessian}. The optimizer-feature configuration adds optimizer control-state
and activation extrema metrics that gradient norms alone do not
capture.
We record the two configurations separately and treat them as
alternative runtime feature designs (Section~\ref{sec:schemas}).

\textbf{Scope.}
We use \dataset to measure how evaluation strategy and runtime feature set jointly determine diagnostic performance and generalizability (Table~\ref{tab:dataset}). Mutation injection gives every trace a known label and a reproducible fault condition. This controlled design isolates evaluation bias and limits our conclusions to the injected-fault corpus.
% --- Table: Corpus composition ---
\begin{table}[htbp]
  \centering\footnotesize
  \caption{Corpus composition of \dataset}
  \label{tab:dataset}
  \setlength{\tabcolsep}{6pt}
  \begin{tabular}{@{}lrl@{}}
    \toprule
    \headrow
    \textbf{Item} & \textbf{Count} & \textbf{Notes} \\
    \midrule
    Total trace files   & 5{,}542 & One CSV per mutated run \\
    Non-empty runs      & 4{,}980 & At least one epoch row recorded \\
    Crash-before-epoch  &   562 & Header only, no epoch data \\
    Distinct programs   &    38 & Unique program identifiers \\
    \CF\ runs           & 4{,}175 & Curvature feature set \\
    \OF\ runs           & 1{,}367 & Optimizer feature set \\
    Unstable runs       & 219 & Loss $>10^6$, 4.4\% of non-empty \\
    \midrule
    \rowcolor{tabhead}
    \multicolumn{3}{@{}l}{\textbf{Fault categories} (runs / programs)} \\
    \midrule
    Activation        &    90 & 2 programs \\
    Hyperparameter    & 1{,}357 & 22 programs \\
    Layer             & 2{,}802 & 38 programs \\
    Loss function     &   991 & 14 programs \\
    Optimization      &   267 & 14 programs \\
    Regularization    &    35 & 1 program \\
    \bottomrule
  \end{tabular}\\[2pt]
  \parbox{\columnwidth}{\scriptsize\raggedright
  \CF\ = curvature feature set (HVP-based), \OF\ = optimizer \& activation feature set.}
\end{table}

\subsection{Two Runtime Feature Sets}
\label{sec:schemas}

The runtime feature configuration determines what a trace contains and
which failure modes can be diagnosed. Adding HVP logging or extreme activation recording can make a previously trained diagnostic model inapplicable, since the model expects a different set of input features. We study two configurations present in \dataset.

\textbf{Curvature features.}
Curvature proxies are derived from the Hessian of the loss with respect
to model parameters,
\begin{equation}
\label{eq:hessian}
  H \;=\; \nabla^{2}_{\theta}\,\mathcal{L}(\theta)
\end{equation}
where $\mathcal{L}$ is the training loss and $\theta$ denotes the
parameter vector.
Forming $H$ explicitly is prohibitive for large networks, so
practical implementations use Pearlmutter's
$R$-operator~\cite{pearlmutter1994}, a differential operator that
propagates directional derivatives through the computation graph.
The $R$-operator computes HVP in $\mathcal{O}(|\theta|)$ time via one additional forward-backward pass.
\begin{equation}
\label{eq:hvp}
  Hv \;=\; \nabla_{\theta}\!\bigl[(\nabla_{\theta}\mathcal{L})^{\!\top} v\bigr]
\end{equation}
From the per-layer HVP vectors we extract scalar summaries, the mean
$\mu_{\mathrm{HVP}}$ and standard deviation $\sigma_{\mathrm{HVP}}$,
together with per-layer HVP norms $\|Hv\|_{2}$.
These curvature metrics are used in curvature-aware
optimization~\cite{martens2010} and have been linked to training
instability at the edge of stability, where the largest
Hessian eigenvalue hovers near
$2/\eta$~\cite{cohen2021edge,arora2022edge,gilmer2022curvature}.
Prior work shows that they localize fault sources in attention-based
architectures more effectively than gradient norms
alone~\cite{jahan2025hessian}.

\textbf{Optimizer features.}
Optimizer features capture first-order gradient statistics and
control-state metrics. Global gradient norm,
\begin{equation}
\label{eq:grad_norm}
  \|\nabla_{\theta}\mathcal{L}\|_{2} \;=\;
  \sqrt{\sum_{i} \bigl(\tfrac{\partial \mathcal{L}}{\partial \theta_i}\bigr)^{2}}
\end{equation}
summarizes the overall magnitude of the update at each training step.
In addition, we record the current learning rate $\eta_{t}$,
per-layer activation statistics (mean $\mu_{\mathrm{act}}$,
standard deviation $\sigma_{\mathrm{act}}$), gradient extrema
($\max|\nabla_{\theta}\mathcal{L}|$, $\min|\nabla_{\theta}\mathcal{L}|$),
and system resource metrics.
These metrics are available in current training debuggers and
monitoring dashboards~\cite{schneider2021cockpit,schoop2021umlaut}.
Table~\ref{tab:runtime_metrics} lists the runtime metrics recorded
under each configuration. Section~\ref{sec:features} describes the
 summary statistics derived from each metric.

\begin{figure*}[htbp]
  \centering
  \resizebox{\textwidth}{!}{%
\begin{tikzpicture}[font=\small]
  \definecolor{pA}{HTML}{5B7BA6}
  \definecolor{pB}{HTML}{C9A24B}
  \definecolor{pC}{HTML}{6FA08C}
  \definecolor{pD}{HTML}{27374D}
  \definecolor{soft}{HTML}{EEF1F6}
  \tikzset{
    stage/.style={draw=gray!50, rounded corners=2.5pt, fill=white, align=center, inner sep=5pt},
    keyst/.style={draw=accent, line width=0.7pt, rounded corners=2.5pt, fill=soft!55, inner sep=4pt},
    phase/.style={font=\scriptsize\sffamily\bfseries, text=gray!70},
    flow/.style={-{Stealth[length=4.5pt,width=3.5pt]}, draw=accent, line width=0.8pt, shorten >=2pt, shorten <=2pt},
    mini/.style={font=\fontsize{6.8}{7.6}\selectfont},
    leg/.style={font=\fontsize{6.0}{6.6}\selectfont, text=black!55},
    hdr/.style={font=\fontsize{6.2}{6.8}\selectfont\sffamily, text=gray!75},
  }
  \def\dx{2.7mm}\def\dr{1.0mm}
  % ===== Stage 1: corpus =====
  \node[stage, minimum width=22mm, minimum height=20mm] (corpus) at (0,0)
    {{\footnotesize\bfseries\color{accent}\textsc{DynFault}}\\[3pt]{\scriptsize\color{black!62}5{,}542 traces}\\[-1pt]{\scriptsize\color{black!62}38 programs}};
  % ===== Stage 2: design =====
  \node[stage, minimum width=25mm, minimum height=20mm] (design) at (3.4,0)
    {{\footnotesize\bfseries\color{accent}Study design}\\[3pt]{\scriptsize\color{black!62}3 diagnostic tasks}\\[-1pt]{\scriptsize\color{black!62}2 feature sets (\textsc{cf},\,\textsc{of})}};
  % ===== Stage 3: EVALUATION (content first, box fitted around it) =====
  \coordinate (ec) at (10.2,0);
  \node[font=\footnotesize\bfseries, text=accent] (etitle) at ([yshift=12.5mm]ec) {Two evaluation splits};
  \node[mini, anchor=west] (wlab) at ([xshift=-27mm,yshift=4.2mm]ec) {Within-program};
  \node[mini, text=accent, anchor=west] (hlab) at ([xshift=-27mm,yshift=-5.0mm]ec) {Program-held-out};
  \coordinate (o)  at ([xshift=0mm,yshift=4.2mm]ec);   % first within-row dot
  \coordinate (o2) at ([xshift=0mm,yshift=-5.0mm]ec);  % first held-out-row dot
  % shaded test-fold column behind both rows
  \fill[accent!13, rounded corners=1pt]
     ([xshift=6*\dx-1.6mm, yshift=2.7mm]o) rectangle ([xshift=7*\dx+1.6mm, yshift=-12.3mm]o);
  % column headers
  \node[hdr] (htr)  at ([xshift=2.5*\dx,yshift=3.6mm]o) {train};
  \node[hdr] (hte)  at ([xshift=6.5*\dx,yshift=3.6mm]o) {test fold};
  % within row dots (random split: interleaved; test reuses programs)
  \foreach \i/\c in {0/pA,1/pB,2/pC,3/pA,4/pB,5/pC,6/pB,7/pA}
    {\fill[\c] ([xshift=\i*\dx]o) circle (\dr);}
  % held-out row dots (grouped; test = unseen program)
  \foreach \i/\c in {0/pA,1/pA,2/pB,3/pB,4/pC,5/pC,6/pD,7/pD}
    {\fill[\c] ([xshift=\i*\dx]o2) circle (\dr);}
  \node[leg, text=accent, anchor=west] (unseen) at ([xshift=7*\dx+2.4mm]o2) {unseen};
  \node[leg] (eleg) at ([yshift=-11.0mm]ec) {each dot = a run \,$\cdot$\, each colour = a program};
  \begin{scope}[on background layer]
    \node[keyst, fit=(etitle)(wlab)(hlab)(htr)(hte)(unseen)(eleg)] (eval) {};
  \end{scope}
  % ===== Stage 4: RESULT (content first, box fitted around it) =====
  \coordinate (fc) at (16.9,0);
  \node[font=\footnotesize\bfseries, text=accent] (ftitle) at ([yshift=12.5mm]fc) {Key finding};
  \coordinate (bz) at ([xshift=-7.5mm,yshift=-9mm]fc);
  \draw[gray!40, line width=0.4pt] ([xshift=-1mm]bz) -- ([xshift=17mm]bz);
  \fill[accent!40] ([xshift=1.5mm]bz) rectangle ++(3.6mm,12.0mm);
  \fill[accent]    ([xshift=11mm]bz) rectangle ++(3.6mm,7.2mm);
  \draw[gray!70, dashed, line width=0.5pt] ([xshift=-1mm,yshift=6.4mm]bz) -- ([xshift=17mm,yshift=6.4mm]bz);
  \node[leg, anchor=west] (chance) at ([xshift=17.4mm,yshift=6.4mm]bz) {chance};
  \node[leg, anchor=north] (lwi) at ([xshift=3.3mm,yshift=-0.6mm]bz) {within};
  \node[leg, anchor=north] (lho) at ([xshift=12.8mm,yshift=-0.6mm]bz) {held-out};
  \coordinate (gx) at ([xshift=15.8mm]bz);
  \draw[gray!45, dashed, line width=0.3pt] ([xshift=5.1mm,yshift=12.0mm]bz) -- ([yshift=12.0mm]gx);
  \draw[gray!45, dashed, line width=0.3pt] ([xshift=14.6mm,yshift=7.2mm]bz) -- ([yshift=7.2mm]gx);
  \draw[{Stealth[length=3pt]}-{Stealth[length=3pt]}, draw=accent, line width=0.5pt] ([yshift=7.2mm]gx) -- ([yshift=12.0mm]gx);
  \node[font=\fontsize{6.6}{7}\selectfont\itshape, text=accent, anchor=west] (gaplab) at ([xshift=0.5mm,yshift=9.6mm]gx) {gap};
  \node[leg, rotate=90] (axlab) at ([xshift=-2.0mm,yshift=3mm]bz) {balanced accuracy};
  \begin{scope}[on background layer]
    \node[stage, fit=(ftitle)(axlab)(chance)(lwi)(lho)(gaplab)] (find) {};
  \end{scope}
  % ===== flow arrows =====
  \draw[flow] (corpus.east) -- (design.west);
  \draw[flow] (design.east) -- (design.east -| eval.west);
  \draw[flow] (eval.east |- design.east) -- (find.west |- design.east);
  % ===== phase labels =====
  \node[phase, above=3pt of corpus.north] {DATA};
  \node[phase, above=3pt of design.north] {SETUP};
  \node[phase, above=3pt of eval.north] {EVALUATION};
  \node[phase, above=3pt of find.north] {RESULT};
\end{tikzpicture}%
  }
  \caption{Study design overview}
  \label{fig:overview}
\end{figure*}
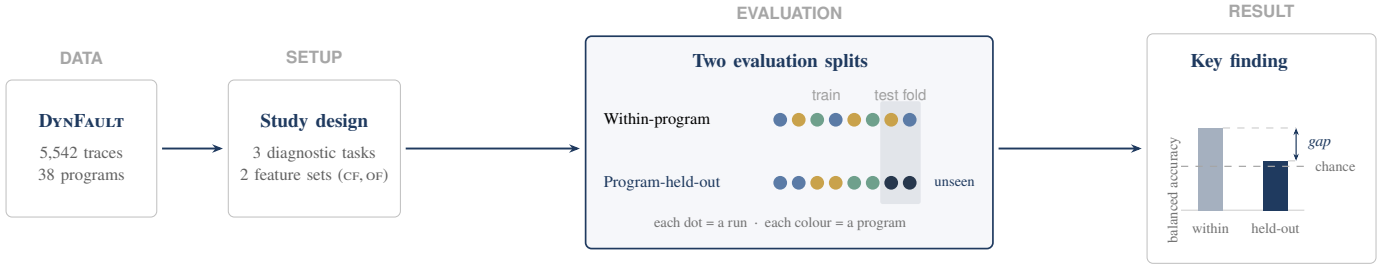

% --- Table: Runtime metrics ---
\begin{table}[htbp]
  \centering\footnotesize
  \caption{Runtime metrics per feature configuration}
  \label{tab:runtime_metrics}
  \setlength{\tabcolsep}{4pt}
  \resizebox{\columnwidth}{!}{%
  \begin{tabular}{@{}llcc@{}}
    \toprule
    \headrow
    \textbf{Group} & \textbf{Metric}
      & \textbf{\CF} & \textbf{\OF} \\
    \midrule
    \multirow{4}{*}{Performance}
      & Training loss 
        & $\checkmark$ & $\checkmark$ \\
      & Validation loss 
        & $\checkmark$ & $\checkmark$ \\
      & Training accuracy
        & $\checkmark$ & $\checkmark$ \\
      & Validation accuracy 
        & $\checkmark$ & $\checkmark$ \\
    \midrule
    \multirow{2}{*}{Curvature}
      & Mean HVP ($\mu_{\mathrm{HVP}}$)
        & $\checkmark$ & --- \\
      & HVP standard deviation ($\sigma_{\mathrm{HVP}}$)
        & $\checkmark$ & --- \\
    \midrule
    \multirow{3}{*}{Gradient}
      & Mean gradient 
        & --- & $\checkmark$ \\
      & Gradient standard deviation
        & --- & $\checkmark$ \\
      & Gradient extrema (min, median, max)
        & --- & $\checkmark$ \\
    \midrule
    \multirow{3}{*}{Optimizer / activation}
      & Learning rate ($\eta_t$)
        & --- & $\checkmark$ \\
      & Activation mean ($\mu_{\mathrm{act}}$)
        & --- & $\checkmark$ \\
      & Activation standard deviation ($\sigma_{\mathrm{act}}$)
        & --- & $\checkmark$ \\
    \midrule
    \multirow{3}{*}{System resources}
      & CPU usage
        & --- & $\checkmark$ \\
      & GPU memory usage
        & --- & $\checkmark$ \\
      & System memory usage
        & --- & $\checkmark$ \\
    \bottomrule
  \end{tabular}%
  }\\[2pt]
  \parbox{\columnwidth}{\scriptsize\raggedright
  $\checkmark$ recorded, --- not recorded.
  \CF: 6 metrics (4 performance + 2 curvature). \OF: 15 metrics (4 performance + 11 optimizer and system).
  HVP = Hessian-vector product.}
\end{table}

Table~\ref{tab:runtime_metrics} lists the global runtime channels shared across the two feature configurations. The instability analyses additionally include per-layer and profile summaries derived from these channels. The feature counts in Table~\ref{tab:instability_models} therefore exceed the channel counts shown here.
% --------------------------------------------------------------
% -----------------------------------------------------------------------
\section{Methodology}
\label{sec:method}

We outline the diagnostic tasks, feature extraction procedures, and evaluation protocols in Fig.~\ref{fig:overview}.

\subsection{Analysis Subsets}
\label{sec:subsets}

All learning-based analyses require epoch-level metrics.
We therefore restrict to \emph{non-empty} traces (Table~\ref{tab:dataset})
that contain at least one logged epoch row. The 562 crash traces fall predominantly in Layer (355) and Hyperparameter (131), the two largest categories, with the rest in Loss (46), Optimization (19), and Activation (11). Their exclusion does not appear to materially distort the fault-category distribution.
Some analyses use configuration-specific subsets.
Curvature-feature (\CF) and optimizer-feature (\OF) traces were collected
for different sets of programs.
When comparing absolute performance across \CF and \OF, we either
(i) restrict to the subset of programs that appear under both feature
configurations (overlap-controlled comparison), or (ii) compare
\emph{within-configuration} generalizability gaps under the same feature
configuration.

\subsection{Diagnostic Tasks}
\label{sec:tasks}

\subsubsection{Fault-type classification}
We predict the mutation category of a trace at three granularities:
(i) six-class fault type, (ii) four-class core fault type, and
(iii) three-class fault family.
This task is the primary subject of RQ1.
Program-held-out evaluation requires training examples for each
class in every fold, so we treat the four-class task as the primary
cross-program estimate (Section~\ref{sec:rq1}).

\subsubsection{Catastrophic instability triage}
We label a trace as \emph{unstable} if the training or validation loss
exceeds $10^6$ at any logged epoch.
This threshold captures unambiguous numerical divergence.
Under standard cross-entropy losses, a value of $10^6$ corresponds to predicting
near-zero probability for the true class across batches, a condition
that is generally not recoverable in practice.
A sensitivity check at thresholds $10^3$ and $10^4$ shows that the
epoch-0 concentration result holds (Section~\ref{sec:rq2}).

\subsubsection{Training/validation performance mismatch detection (\tvmm)}
We label a stable trace as a \tvmm case if the per-epoch boolean flag
\texttt{acc\_gap\_too\_big} is true in at least one epoch of the observed
window.
The flag is set at each epoch $e$ when the absolute
training--validation accuracy gap meets or exceeds 0.1
(10\%):
\begin{equation}
\texttt{acc\_gap\_too\_big}_{e}
  \;=\;
  1\!\bigl[|\mathrm{acc}_{\mathrm{train},e}
                    - \mathrm{acc}_{\mathrm{val},e}|
                    \;\geq\; 0.1\bigr].
\label{eq:tvmm}
\end{equation}
This threshold and computation rule originate from the DEFault corpus
construction protocol~\cite{jahan2025default} and are applied
identically (see our replication package~\cite{icsmeReplicationPackage}).
To isolate mismatch effects from catastrophic failures, we restrict this
task to stable traces only (i.e., no loss explosion in any epoch).

\subsection{Feature Extraction}
\label{sec:features}

For each scalar metric channel $z_e$ observed over the first $k$ logged
epochs $e \in \{0,\dots,k-1\}$, we extract two summary statistics: the
window mean $\bar{z}_k$ and the ordinary least-squares slope
$\hat{\beta}_k$,
\begin{equation}
\bar{z}_k = \frac{1}{k}\sum_{e=0}^{k-1} z_e\,,\qquad
\hat{\beta}_k = \frac{\sum_{e=0}^{k-1}(e - \bar{e})\,(z_e - \bar{z}_k)}
                     {\sum_{e=0}^{k-1}(e - \bar{e})^2}\,,
\label{eq:features}
\end{equation}
where $\bar{e} = (k-1)/2$.
For a trace with $d$ scalar channels, the feature vector at observation window
$k$ is $\mathbf{x}_k \in \mathbb{R}^{2d}$ formed by concatenating the
mean and slope for each channel.
When $k{=}1$, the slope term is undefined, so we use only the
mean statistics at $k{=}1$, giving $d$ features. We consider observation windows $k \in \{1, 3, 5, 10, 20, 50\}$.
We standardize features using statistics from the training split only
and clip values to $[\pm 10^{12}]$.
Missing channels are imputed with column medians computed on the
training split. 

We use logistic regression with class-balanced weights and a one-vs-rest formulation for multi-class tasks. This setting provides a controlled linear probe of the feature representation under each evaluation protocol, keeping classifier capacity fixed while comparing within-program and program-held-out splits. Although logistic regression is linear, it can still capture program-level differences when those differences are present in the features. We use it to test whether the evaluation gap appears even without a high-capacity classifier. For comparisons with existing fault diagnosis techniques~\cite{jahan2025default,cao2022deepfd}, we report their original Random Forest classifiers alongside logistic regression (Section~\ref{sec:comparison}). On the same DEFault feature set, the gap widens from 0.088 under logistic regression to 0.34 under random forest (Fig.~\ref{fig:comparison_ba_gap}), so the logistic-regression gap we report for RQ1 is the conservative estimate. Logistic regression also enables the per-fold coefficient analysis in Fig.~\ref{fig:csi_coeff} that a random-forest ensemble would mask.

\subsection{Evaluation Strategies}
\label{sec:strategies}
\subsubsection{Within-program evaluation}
We perform stratified $K$-fold cross-validation over individual runs ($K{=}5$). Runs from the same program may appear in both training and test folds. This protocol measures \emph{within-program reuse}, where a trained diagnostic model is applied to new runs of a program already represented in
training data.

\subsubsection{Program-held-out evaluation}
We perform grouped $K$-fold cross-validation over runs ($K{=}5$) with program identifiers as the grouping variable, so that each test fold contains only programs
entirely absent from the corresponding training fold.
This protocol measures \emph{cross-program deployment}, where a model is applied
to a program not seen during training. For the cluster-based curvature triage heuristic (Algorithm~\ref{alg:triage}),
we additionally report a stricter leave-one-program-out (LOPO) evaluation
that holds out each program in turn. We use LOPO because the heuristic is trained
without labels and produces a single interpretable ``terminate'' action,
for which per-program precision is directly interpretable.

\subsubsection{Metrics and uncertainty}
For multi-class tasks we report balanced accuracy. Unless otherwise noted, \textit{``accuracy'' refers to balanced accuracy throughout this paper}.
For binary tasks we report \ROCAUC and \PRAUC. PR and ROC curves pool out-of-fold
predicted probabilities across folds.
We compute 95\% confidence intervals via a cluster bootstrap over programs
(with replacement)~\cite{cameron2008bootstrap}, preserving within-program dependence.

\subsubsection{Evaluation-strategy gap}
Let $M_{\mathrm{within}}(k)$ and $M_{\mathrm{heldout}}(k)$ denote the
value of a metric~$M$ obtained under within-program and program-held-out
evaluation, respectively, for the same classifier and feature vector
$\mathbf{x}_k$.
The \emph{evaluation-strategy gap} is
\begin{equation}
\Delta_M(k) \;=\; M_{\mathrm{within}}(k) \;-\; M_{\mathrm{heldout}}(k)\,.
\label{eq:gap}
\end{equation}
A positive $\Delta_M(k)$ indicates that within-program evaluation reports
a higher figure than program-held-out evaluation at~$k$ observed epochs.

\subsection{Evaluation-Strategy Controls}
\label{sec:controls}
We use three classifier-based controls to identify the source of the within-program advantage.

\subsubsection{Program-identity prediction}
We predict the program identifier from early-epoch features.
High accuracy indicates that feature vectors encode program-level
structure beyond fault semantics.

\subsubsection{Within-program label permutation}
We randomly permute fault labels \emph{within each program}, preserving
each program's label frequencies, and re-evaluate under within-program
cross-validation. Above-chance performance after permutation indicates that within-program
evaluation can succeed using program-level structure even when fault
labels carry no discriminative information.

\subsubsection{Per-program normalization (analysis-only control)}
As an analysis-only control (i.e., not deployable in practice), we $z$-normalize features within each program
(i.e., subtract the per-program mean and divide by the per-program standard deviation) \emph{before} any cross-validation split, since this uses unlabeled information from the held-out programs.
If within-program performance drops substantially while program-held-out performance changes little, the evaluation gap is driven by
program-level feature statistics rather than transferable fault signal.

\section{Study Findings}
\label{sec:results}

\subsection{Evaluation-Strategy Gap}
\label{sec:rq1}

\smallskip\noindent\textbf{Answering RQ$\mathbf{_1}$: The within-program vs.\ program-held-out gap and its cause}

To answer RQ$_1$, we measure the within-program and program-held-out balanced accuracy for fault-type classification, and then apply four controls to identify what produces any difference between them.

Table~\ref{tab:leakage_k5} reports \BA at \kb{5}, a representative early-epoch window. The full analysis across all $k$ values is in our replication package~\cite{icsmeReplicationPackage}.
We treat the four-class task (Layer, Hyperparameter, Loss, Optimization) as the primary estimate because each class
appears in at least 14 programs, making program-held-out evaluation
well-defined. The six-class task adds Activation (2~programs) and
Regularization (1~program), so some program-held-out folds lack
training signal for those labels. We report it for completeness and
interpret its near-chance cross-program \BA accordingly.
This near-chance result ($0.178 \approx 0.17$ chance) reflects the sparse program coverage for these two categories rather than a genuine absence of transferable signal.

\begin{table}[t]
  \centering\footnotesize
  \caption{Evaluation-strategy gap at \kb{5}}
  \label{tab:leakage_k5}
  \setlength{\tabcolsep}{3pt}
  \resizebox{\columnwidth}{!}{%
  \begin{tabular}{@{}lccrrr@{}}
    \toprule
    \headrow
    \textbf{Task} & \textbf{Classes} & \textbf{Chance}
      & \textbf{Within-prog.}
      & \textbf{Cross-prog.}
      & \textbf{Gap} \\
    \midrule
    \findrow
    Fault type & 4 & 0.25
      & 0.469 \ci{0.343}{0.545}
      & 0.279 \ci{0.232}{0.319}
      & 0.190 \ci{0.053}{0.258} \\
    Fault type   & 6 & 0.17
      & 0.468 \ci{0.310}{0.500}
      & 0.178 \ci{0.160}{0.281}
      & 0.290 \ci{0.091}{0.325} \\
    Fault family       & 3 & 0.33
      & 0.518 \ci{0.446}{0.595}
      & 0.391 \ci{0.346}{0.438}
      & 0.128 \ci{0.043}{0.206} \\
    \bottomrule
  \end{tabular}%
  }\\[2pt]
  \parbox{\columnwidth}{\scriptsize\raggedright
  Shaded row is the primary four-class estimate. Values are balanced accuracy with 95\% bootstrap confidence interval in brackets. Chance $=1/$(number of classes). The six-class row is reported for completeness.}
\end{table}

On the four-class task, within-program \BA is 0.469~$\ci{0.343}{0.545}$
and cross-program \BA is 0.279~$\ci{0.232}{0.319}$, a gap of
0.190~$\ci{0.053}{0.258}$.
In our experiments, the program-held-out score is only slightly higher than the 0.25 chance baseline, and its 95\% CI includes chance-level performance. We therefore interpret this result as limited cross-program fault-type discrimination in this corpus, rather than deployment-ready diagnosis. The main empirical finding is the drop from within-program to program-held-out evaluation. Our control experiments suggest that this drop is partly explained by program-specific patterns in the runtime features.

\begin{figure}[htbp]
  \centering
  \subfloat[Four-class fault type\label{fig:leakage_sweep4}]{%
    \includegraphics[width=0.50\linewidth]{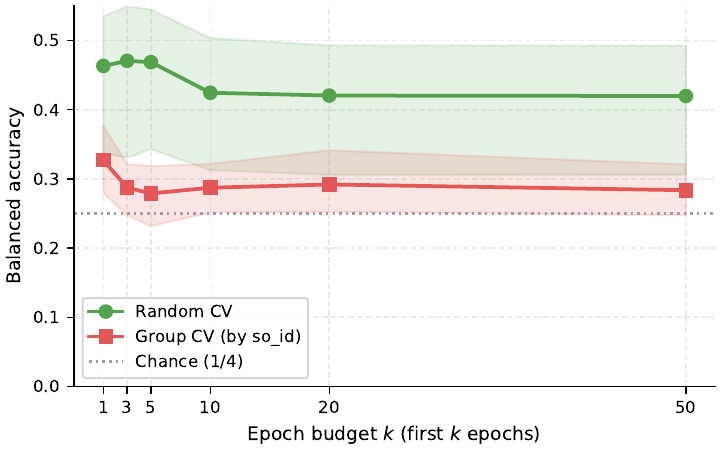}%
  }\hfil
  \subfloat[Six-class fault type\label{fig:leakage_sweep6}]{%
    \includegraphics[width=0.50\linewidth]{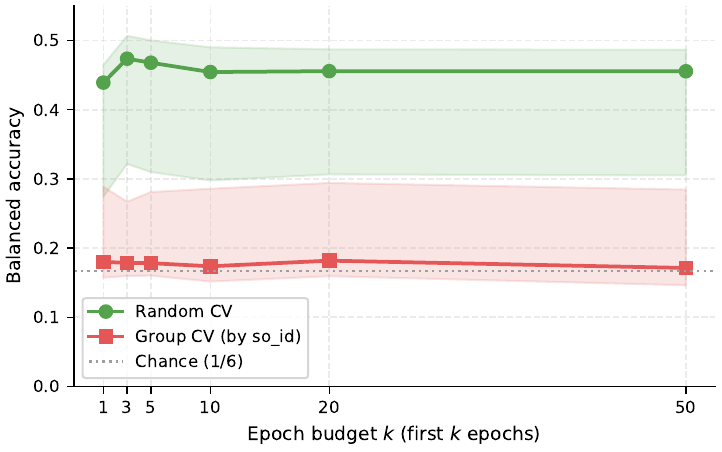}%
  }
  \caption{Evaluation-strategy gap across observation windows $k$} 
  \label{fig:leakage_sweep}
\end{figure}

Fig.~\ref{fig:leakage_sweep} shows that this separation between within-program and program-held-out evaluation holds across all observation windows~$k$ on both task granularities. The primary four-class task preserves a positive gap at every $k$. On the six-class task the gap ranges from approximately 0.13 at $k{=}1$ to 0.29 at $k{=}50$, with the lower bound exceeding zero throughout (the minimum CI lower bound is 0.044, at $k{=}1$), so the gap is not explained by the observation window alone.
We next use three classifier-based controls and one distribution-level analysis to identify its source.

\begin{figure}[htbp]
  \centering
  \subfloat[Predict program identity\label{fig:fingerprint}]{%
    \includegraphics[width=0.50\linewidth]{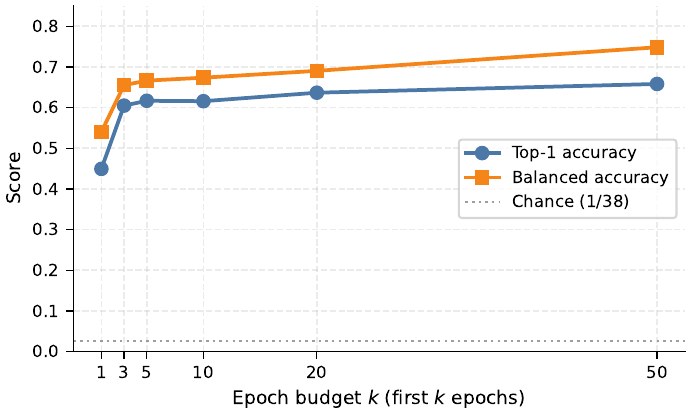}%
  }\hfil
  \subfloat[Within-program \BA\label{fig:perm}]{%
    \includegraphics[width=0.50\linewidth]{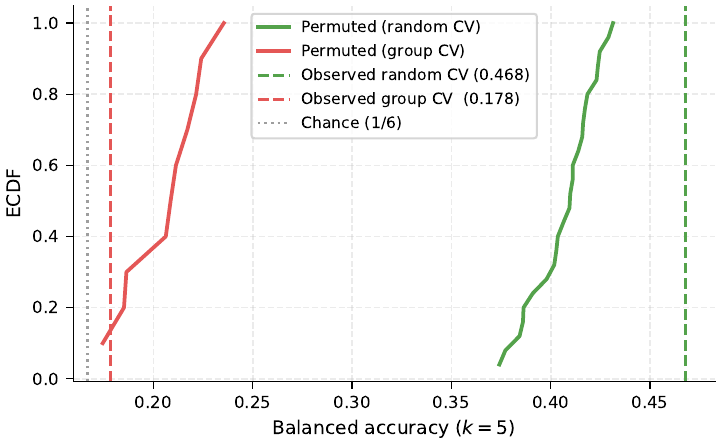}%
  }
  \caption{Program-level structure controls at \kb{5}}
  \label{fig:controls_ab}
\end{figure}

\textit{(a) Program-identity predictability.} Fig.~\ref{fig:fingerprint} shows that 38-class program-identity accuracy reaches 0.62 at \kb{5} (chance is 0.026), rising
monotonically with~$k$. We observed that a diagnostic model operating on these features can identify
the program independently of fault labels, suggesting
that program-level structure contributes to within-program CV
performance.

\textit{(b) Within-program label permutation.} Fig.~\ref{fig:perm} shows that permuting fault labels within each program keeps within-program CV \BA well above chance. Across 25 permutations, median permuted \BA is 0.41, versus a 0.17 chance baseline (6-class uniform) and 0.47 for the unpermuted case. We found that a classifier achieves \BA well above chance
on within-program splits even when fault labels carry no discriminative
information, suggesting that program-level structure in the feature
space contributes to this performance independently of fault-label content.

\textit{(c) Per-program normalization.} Table~\ref{tab:normalization} shows the normalization control.
After $z$-normalizing features within each program, program-identity accuracy drops from 0.617 to 0.096 and within-program fault-type \BA from 0.468 to 0.325, while cross-program \BA remains similar (0.178 to 0.158), indicating that the within-program drop reflects removal of program-level shortcuts rather than loss of generalizable signal.

% --- Table: Normalization control ---
\begin{table}[htbp]
  \centering\footnotesize
  \caption{Per-program normalization control at \kb{5}}
  \label{tab:normalization}
  \setlength{\tabcolsep}{2.5pt}
  \resizebox{\columnwidth}{!}{%
  \begin{tabular}{@{}llrrc@{}}
    \toprule
    \headrow
    \textbf{Metric} & \textbf{Setting} & \textbf{Raw} & \textbf{Normalized} & \textbf{$\Delta$} \\
    \midrule
    \multirow{2}{*}{Fault-type \BA }
      & within & 0.468 \ci{0.310}{0.500}
      & 0.325 \ci{0.229}{0.473} & $-$0.143 \\
      & cross & 0.178 \ci{0.160}{0.281}
      & 0.158 \ci{0.129}{0.269} & $-$0.020 \\
    \addlinespace
    Program-identity accuracy
      & within & 0.617
      & 0.096 & $-$0.521 \\
    \bottomrule
  \end{tabular}%
  }\\[2pt]
  \parbox{\columnwidth}{\scriptsize\raggedright
  Values are balanced accuracy with 95\% confidence interval in brackets (program-identity is plain accuracy). $\Delta$ = normalized $-$ raw.}
\end{table}

\begin{figure}[htbp]
  \centering
  \includegraphics[width=0.90\linewidth]{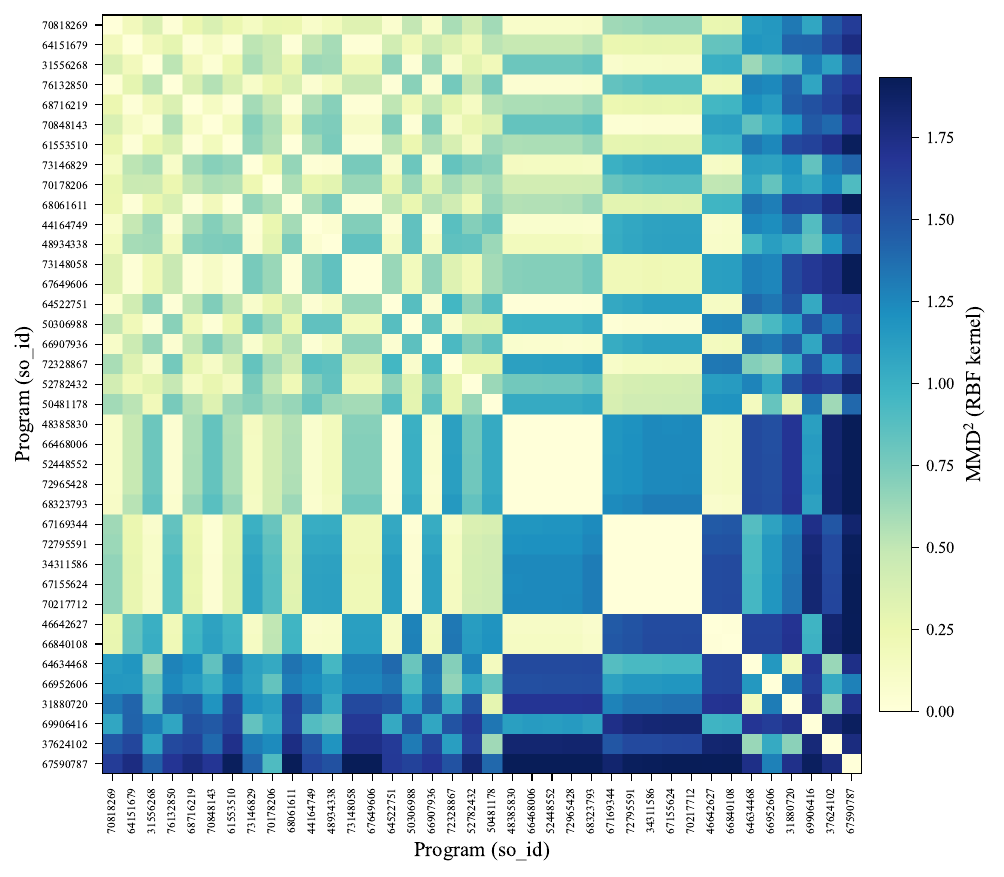}
  \caption{Pairwise MMD$^2$ between program feature distributions at \kb{5}}
  \label{fig:domain_shift}
\end{figure}

\textit{(d) Distribution-level domain shift.}
The controls above use classifier accuracy as a proxy for
program-level structure.
For a distribution-level analysis, we compute the
pairwise Maximum Mean Discrepancy (MMD, RBF kernel with median
heuristic bandwidth)~\cite{gretton2012kernel} between the feature distributions of all
38~programs at \kb{5} using the four shared performance channels
(loss and accuracy, mean and slope).
The median pairwise MMD$^2$ across all 703 program pairs is 0.716,
with a mean of 0.774 and a range of 0.0 to 1.93.
Under the RBF kernel with median heuristic bandwidth, MMD$^2$ is 0 for identical distributions and increases without a fixed upper bound as distributions diverge. A value of 0.716 therefore reflects substantial inter-program separation.
Fig.~\ref{fig:domain_shift} shows the heatmap sorted by
mean domain shift.
The large inter-program distances suggest that programs occupy
distinct feature-space regions, consistent with the program-identity prediction in control~(a).

All four controls point to the same explanation. The within-program CV advantage is attributable to program-level structure in the feature space rather than transferable fault semantics.
This structure is visible at the distribution level (median pairwise
MMD$^2$ 0.716), predictable by a classifier (62\% program-identity
accuracy), exploitable even under permuted labels (\BA 0.41 vs.\
chance 0.17), and removable by per-program normalization without
affecting cross-program performance.
Within-program evaluation remains valid as a measure of reuse performance, and the two strategies capture different signals and should be interpreted accordingly. The remaining questions examine which runtime features and diagnostic tasks remain effective under the harder cross-program standard.

\begin{rqbox}
\textbf{RQ$\mathbf{_1}$ Summary.}
We found that within-program fault-type \BA exceeds cross-program \BA by 0.190 at \kb{5}. Four converging controls confirm that program-level feature structure, rather than fault signal, accounts for this gap. These results show that within-program evaluation substantially overstates diagnostic ability.
\end{rqbox}

\subsection{Runtime Feature Generalizability}
\label{sec:rq_features}

\smallskip\noindent\textbf{Answering RQ$\mathbf{_2}$: Instability timing and curvature-based detection on unseen programs}
\label{sec:rq2}

To answer RQ$_2$, we first determine when catastrophic instability occurs during training, and then test whether adding curvature features improves its detection on unseen programs.

\begin{figure}[htbp]
  \centering
  \subfloat[Timing of first loss explosion\label{fig:instability_timing}]{%
    \includegraphics[width=0.50\linewidth]{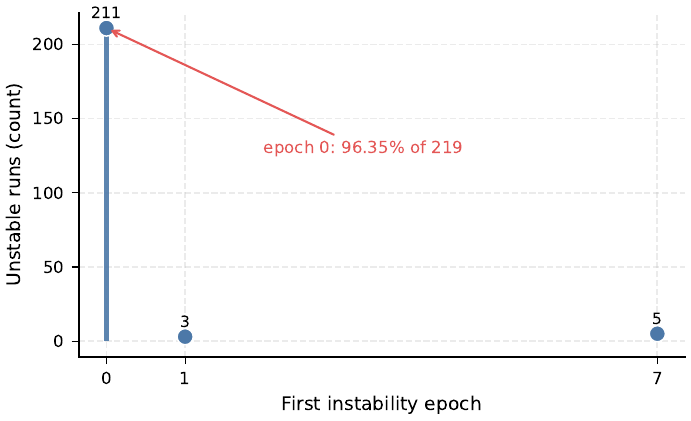}%
  }\hfil
  \subfloat[Curvature feature benefit\label{fig:instability_auc}]{%
    \includegraphics[width=0.50\linewidth]{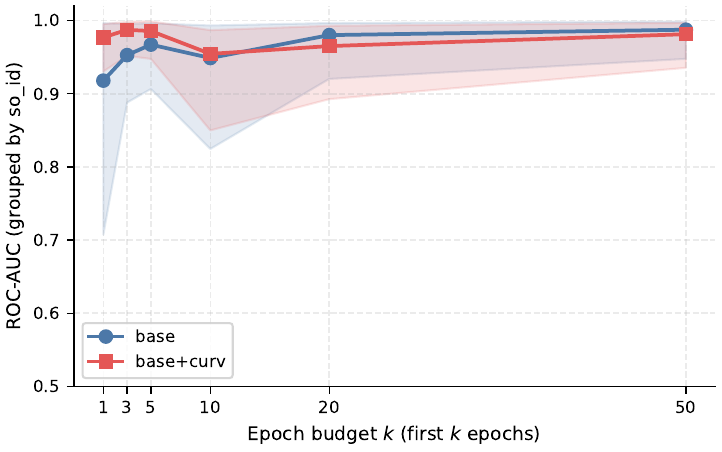}%
  }
  \caption{Instability characterization (a)~96\% of unstable runs explode at epoch~0 (b)~adding curvature features improves cross-program detection at every~$k$}
  \label{fig:instability_combined}
\end{figure}

\textit{Instability timing.} Fig.~\ref{fig:instability_timing} shows the distribution of the
first epoch of loss explosion across all 219 unstable runs. Of these, 96\%
(211 of 219) explode at epoch~0 and 98\% (214) by epoch~1.
We found that this concentration holds across the instability threshold.
At thresholds $10^3$ and $10^4$, 99.0\% and 91.9\% of unstable runs,
respectively, first explode at epoch~0 (vs.\ 96.3\% at $10^6$).
Lowering the threshold increases the flagged set (11.9\% of runs at $10^3$
and 6.9\% at $10^4$, vs.\ 4.4\% at $10^6$), so $10^6$ is the most
conservative threshold.
These results indicate that, in our dataset, instability monitoring is most informative at initialization, before substantial training compute is spent.

\textit{Curvature feature addition.}
Table~\ref{tab:instability_models} reports instability detection
under program-held-out evaluation for baseline features and baseline
combined with curvature features. At \kb{1}, adding curvature features improves \ROCAUC from
0.918~$\ci{0.707}{0.996}$ to 0.977~$\ci{0.930}{0.995}$.
The loss-ablated variants suggest that this
gain is not explained solely by rising loss values. In our study, curvature provided an additional early indicator of instability.
\PRAUC confidence intervals are wide at the 4.4\% positive rate,
because cluster-bootstrap resampling can produce folds with very few
unstable runs, so \ROCAUC is the primary metric.
Figure~\ref{fig:instability_auc} shows curvature benefit across all~$k$.

% --- Table: Instability models ---
\begin{table}[htbp]
  \centering
  \footnotesize
  \caption{Cross-program instability detection}
  \label{tab:instability_models}
  \setlength{\tabcolsep}{4pt}
  \resizebox{\columnwidth}{!}{%
  \begin{tabular}{@{}llrll@{}}
    \toprule
    \headrow
    \textbf{Feature set} & $k$ & \textbf{\#Features}
      & \textbf{\ROCAUC}
      & \textbf{\PRAUC} \\
    \midrule
    Baseline          & 1 & 19 & 0.918 \ci{0.707}{0.996} & 0.776 \ci{0.466}{0.956} \\
    Baseline          & 5 & 38 & 0.967 \ci{0.906}{0.996} & 0.768 \ci{0.635}{0.954} \\
    \addlinespace
    Baseline $+$ \CF  & 1 & 22 & 0.977 \ci{0.930}{0.995} & 0.896 \ci{0.700}{0.977} \\
    Baseline $+$ \CF  & 5 & 43 & 0.986 \ci{0.948}{0.999} & 0.894 \ci{0.752}{0.991} \\
    \addlinespace
    No-loss baseline  & 1 & 17 & 0.850 \ci{0.566}{0.966} & 0.547 \ci{0.181}{0.911} \\
    No-loss baseline  & 5 & 34 & 0.954 \ci{0.888}{0.988} & 0.601 \ci{0.345}{0.899} \\
    \addlinespace
    No-loss $+$ \CF   & 1 & 20 & 0.896 \ci{0.743}{0.966} & 0.756 \ci{0.312}{0.916} \\
    No-loss $+$ \CF   & 5 & 39 & 0.950 \ci{0.865}{0.980} & 0.684 \ci{0.325}{0.930} \\
    \bottomrule
  \end{tabular}%
  }\\[2pt]
  \parbox{\columnwidth}{\scriptsize\raggedright
  Program-held-out evaluation, values with 95\% confidence interval in brackets. ROC-AUC and PR-AUC = area under the ROC \& precision--recall curves.}
\end{table}

\begin{figure}[htbp]
  \centering
  \subfloat[Mean HVP quintiles\label{fig:quintile_mean_hvp}]{%
    \includegraphics[width=0.50\linewidth]{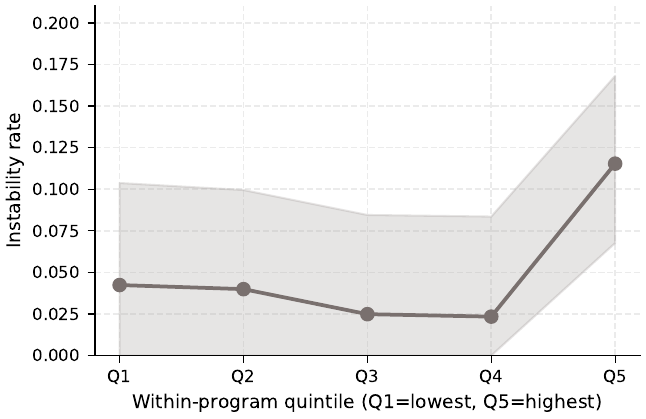}%
  }\hfil
  \subfloat[Gradient std quintiles\label{fig:quintile_grad_std}]{%
    \includegraphics[width=0.50\linewidth]{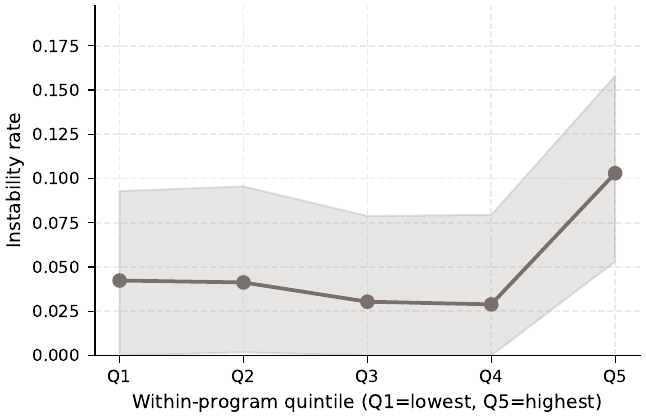}%
  }
  \caption{Instability rate by epoch-0 quintile}
  \label{fig:quintiles}
\end{figure}

\begin{figure}[htbp]
  \centering
  \includegraphics[width=0.90\linewidth]{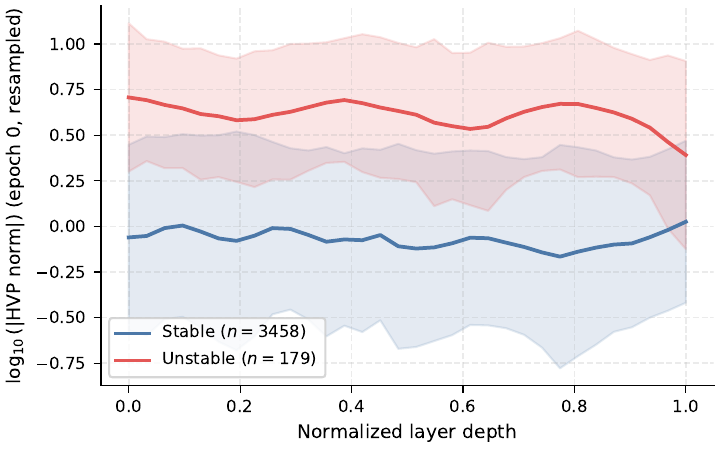}
  \caption{Layer-wise HVP magnitude at epoch~0}
  \label{fig:hvp_profile}
\end{figure}

\textit{Cross-program risk stratification.}
Fig.~\ref{fig:quintiles} reports instability prevalence by quintile
of mean HVP and gradient standard deviation at epoch-0.
We observed that both metrics exhibit a consistent risk gradient across quintiles, which suggests that they may support a cross-program triage heuristic without a trained classifier. Fig.~\ref{fig:hvp_profile} shows the underlying pattern, where unstable runs exhibit uniformly elevated HVP magnitude across all
network layers (median $\approx$0.7 log units above stable runs),
suggesting that loss-surface curvature at initialization may be a
whole-network property rather than a layer-specific artifact.

% --- Table: Curvature cluster types ---
\begin{table}[htbp]
  \centering
  \footnotesize
  \caption{Curvature cluster patterns}
  \label{tab:phenotypes}
  \setlength{\tabcolsep}{2.5pt}
  \resizebox{\columnwidth}{!}{%
  \begin{tabular}{@{}crrrrrr@{}}
    \toprule
    \headrow
    \textbf{Cluster} & $n$ & \textbf{Programs}
      & \textbf{Unstable (\%)}
      & \textbf{Mismatch (\%)}
      & \textbf{Median HVP}
      & \textbf{Median $\nabla$-std} \\
    \midrule
    0 & 1346 & 19 &  0.5 & 49.1 & $6.1\times10^{-1}$ & $1.4\times10^{-2}$ \\
    \rowcolor{rowhi}
    1 &  121 &  9 & \textbf{100.0} &  3.3 & $5.4\times10^{6}$ & $6.8\times10^{5}$ \\
    2 &  367 &  8 &  0.8 & 28.6 & $6.4\times10^{-1}$ & $8.8\times10^{-2}$ \\
    3 &  100 &  4 & 10.0 &  0.0 & $9.2\times10^{3}$  & $1.0\times10^{4}$  \\
    4 &  536 &  6 &  2.2 &  \textbf{0.9} & $1.9\times10^{-1}$ & $1.5\times10^{-2}$ \\
    5 &  933 & 17 &  1.9 & 50.2 & $5.2\times10^{-2}$ & $1.4\times10^{-2}$ \\
    6 &  234 &  8 &  3.4 &  8.5 & $4.0\times10^{1}$  & $1.8\times10^{1}$  \\
    \bottomrule
  \end{tabular}%
  }\\[2pt]
  \parbox{\columnwidth}{\scriptsize\raggedright
  $K{=}7$ partition of 3{,}637 epoch-0 curvature-feature runs. $n$ = runs in cluster, $\nabla$-std = gradient standard deviation, HVP = Hessian-vector product. Low-curvature clusters (C0, C4, C5) show near-zero instability but differ sharply in mismatch prevalence.}
\end{table}

\begin{figure}[htbp]
  \centering
  \includegraphics[width=0.90\linewidth]{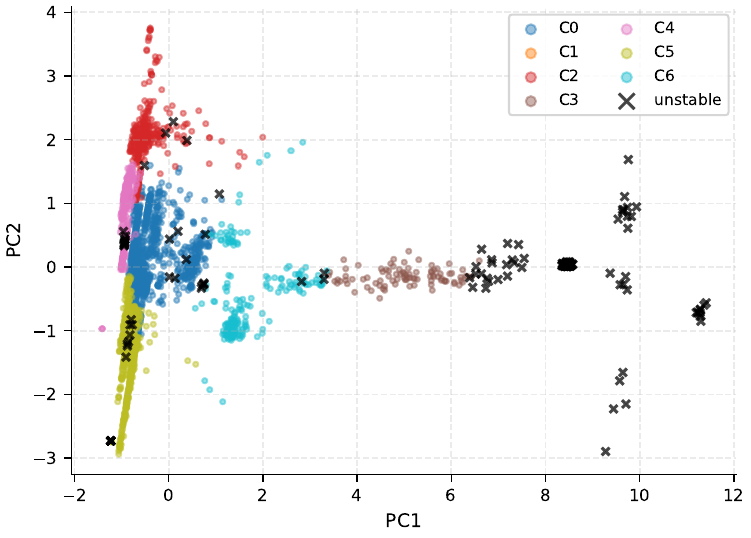}
  \caption{PCA of epoch-0 curvature features by cluster ($K{=}7$)}
  \label{fig:phenopca}
\end{figure}

\textit{Curvature-based triage rule.}
We also examined whether epoch-0 curvature geometry can group runs into interpretable categories before training proceeds.
We cluster the 3,637 curvature-feature runs by epoch-0 curvature
geometry (HVP magnitude, gradient standard deviation, per-layer
profile summaries) using $k$-means.
Average silhouette coefficients peak at
$K{=}2$ (0.742), indicating a dominant split between high-curvature and low-curvature runs. For the phenotype analysis we fit a finer $K{=}7$ partition over all 3,637 curvature-feature runs (Table~\ref{tab:phenotypes}, Fig.~\ref{fig:phenopca}), which resolves outcome variation inside the low-curvature majority while keeping the extreme high-curvature cluster.
The three largest low-curvature clusters (C0, C4, C5) share near-zero instability rates but differ sharply in mismatch prevalence. Table~\ref{tab:phenotypes} summarizes each cluster and Fig.~\ref{fig:phenopca} shows the PCA embedding.

\begin{algorithm}[!htbp]
\footnotesize
\setlength{\algomargin}{4pt}
\caption{LOPO Curvature Triage}
\label{alg:triage}
\KwIn{Epoch-0 \CF\ vectors $\{\mathbf{x}_0^{(i)}\}$,
      program labels $\{p^{(i)}\}$,
      instability labels $\{y^{(i)}\}$,
      clusters $K$}
\KwOut{LOPO precision and recall}
$\mathrm{TP} \leftarrow 0$\;
$\mathrm{FP} \leftarrow 0$\;
$\mathrm{FN} \leftarrow 0$\;
\ForEach{unique program $p \in \mathcal{P}$}{
  $\mathcal{D}_{\mathrm{train}} \leftarrow \{\mathbf{x}_0^{(i)} : p^{(i)} \neq p\}$\;
  $\mathcal{D}_{\mathrm{test}}  \leftarrow \{\mathbf{x}_0^{(i)} : p^{(i)} = p\}$\;
  $\{\boldsymbol{\mu}_1,\dots,\boldsymbol{\mu}_K\} \leftarrow k\text{-means}(\mathcal{D}_{\mathrm{train}},\, K)$\;
  $c^{*} \leftarrow \arg\max_{j}\,\boldsymbol{\mu}_j[\text{mean-HVP}]$
  \tcp*[f]{highest mean-HVP centroid}\;
  \ForEach{$(\mathbf{x}_0^{(i)},\, y^{(i)}) \in \mathcal{D}_{\mathrm{test}}$}{
    $j^{*} \leftarrow \arg\min_{j}\,\lVert \mathbf{x}_0^{(i)} - \boldsymbol{\mu}_j \rVert_2$\;
    \eIf{$j^{*} = c^{*}$}{
      \textsc{terminate}\;
      \lIf{$y^{(i)} = 1$}{$\mathrm{TP} \leftarrow \mathrm{TP} + 1$}
      \lElse{$\mathrm{FP} \leftarrow \mathrm{FP} + 1$}
    }{
      \textsc{continue}\;
      \lIf{$y^{(i)} = 1$}{$\mathrm{FN} \leftarrow \mathrm{FN} + 1$}
    }
  }
}
\Return $\dfrac{\mathrm{TP}}{\mathrm{TP}+\mathrm{FP}}$,~
        $\dfrac{\mathrm{TP}}{\mathrm{TP}+\mathrm{FN}}$\;
\end{algorithm}

\textit{Zero-false-positive early-termination rule.}
Cluster C1 ($n{=}121$, median mean HVP $5.37\!\times\!10^6$,
9~programs) is 100\% unstable.
Treating C1 membership as a termination signal achieves precision
1.00, recall 0.676, and F1 0.807 on the 179 unstable curvature-feature
runs in the full in-sample evaluation. Of these, 121 runs can be
terminated at epoch~0 with zero false positives.

\textit{Evaluation protocol distinction.}
The supervised instability models in
Table~\ref{tab:instability_models} use \emph{program-held-out grouped
5-fold cross-validation} (Section~\ref{sec:strategies}).
The triage heuristic below uses a stricter
\emph{leave-one-program-out} (LOPO) protocol because it is unsupervised
and produces a single binary ``terminate'' action for
which per-program precision is directly interpretable.
Grouped 5-fold provides aggregate metrics with bootstrap confidence intervals. LOPO characterizes per-program precision of the decision rule. We assess out-of-sample precision using the LOPO triage evaluation in
Algorithm~\ref{alg:triage}. Across the 24 programs that have curvature-feature traces (of 38 total), which contain 179 unstable
curvature-feature runs, the cross-program
aggregate achieves precision 1.00 (TP~$=10$, FP~$=0$) and recall
0.056 (10 of 179, FN~$=169$).
We observed that the zero-false-positive property holds out-of-sample, since no stable run
was assigned to a high-HVP cluster in our evaluation.
LOPO recall is low because the extreme-HVP signature
driving C1 (${\sim}5\times10^6$) appears concentrated in a few programs.
When C1-contributing programs are held out, their unstable runs
are assigned to the next-highest-HVP cluster rather than detected.
The rule therefore provides reliable precision across programs but
limited recall outside the programs that formed C1.
Both figures must be reported together. In-sample recall 0.676 and F1 0.807 characterize the rule within a known program subset, and cross-program recall 0.056 characterizes its generalizability.

\begin{figure}[htbp]
  \centering
  \includegraphics[width=0.90\linewidth]
  {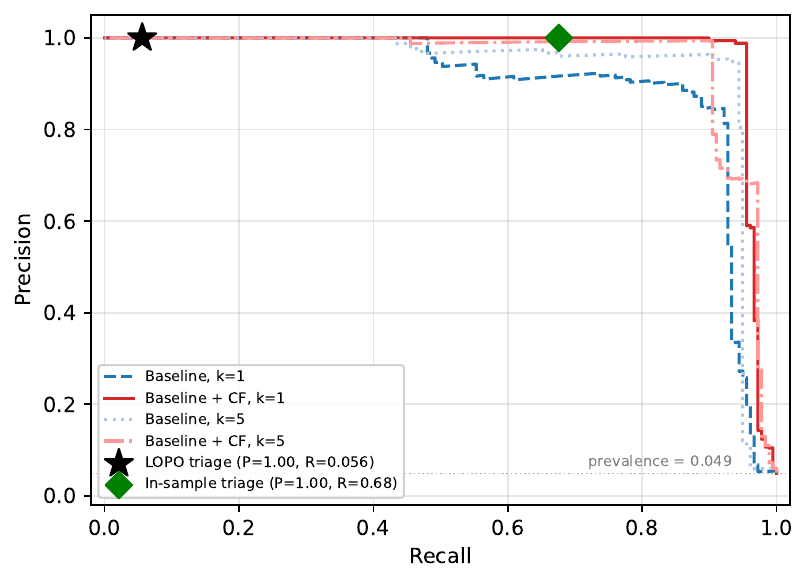}
  \caption{Instability precision--recall curves under program-held-out evaluation}
  \label{fig:operating_points}
\end{figure}

\textit{Precision-recall operating points.}
The low LOPO recall represents one operating point on a
precision-recall trade-off.
Fig.~\ref{fig:operating_points} shows PR curves
under program-held-out evaluation for baseline and
baseline$+$\CF classifiers at \kb{1} and \kb{5}.
At \kb{1}, the baseline$+$\CF classifier achieves 0.90~precision at
0.96~recall, indicating that supervised curvature-based models can achieve both high precision and high recall in this setting.
The LOPO cluster rule (star) operates at the extreme
high-precision end (precision 1.00, recall 0.056),
while the in-sample cluster rule (diamond) lies at precision 1.00
with recall 0.676.
The operating point depends on the relative cost of false positives vs.\ missed instabilities.

\textit{Mismatch-prevalence variation.}
Clusters C0 and C5 ($n{=}1346$ and $n{=}933$) exhibit mismatch
prevalences of 49\% and 50\%, while C4 ($n{=}536$) has 0.9\%.
All three have instability rates below 2.3\%.
This 55$\times$ difference in mismatch prevalence, invisible at
$K{=}2$, suggests that \tvmm monitoring may be needed for some
low-curvature program types but not others.
These cluster patterns suggest that epoch-0 assignment may help prioritize monitoring. Programs resembling C0 or C5 showed high mismatch prevalence, whereas those resembling C4 did not.

\begin{rqbox}
\textbf{RQ$\mathbf{_2}$ Summary.}
We found that most instability originates at initialization, with 96\% of explosive runs occurring at epoch~0. Incorporating curvature-based features improves cross-program detection (\ROCAUC $0.918 \rightarrow 0.977$ at $k{=}1$). The unsupervised LOPO triage rule provides a high-precision but low-recall operating point, while supervised curvature models offer adjustable precision--recall trade-offs.
\end{rqbox}

\smallskip\noindent\textbf{Answering RQ$\mathbf{_3}$: Optimizer features for mismatch detection on unseen programs}
\label{sec:rq3}

To answer RQ$_3$, we compare optimizer features against a loss-and-accuracy baseline for mismatch detection, under both evaluation strategies, on an overlap-controlled program subset and on the full subsets.

\textit{Overlap-controlled comparison (primary).}
The two configurations cover different programs and mismatch
prevalences, so we compare them on an overlap-controlled program subset.
We therefore interpret absolute \CF-vs-\OF differences cautiously and focus on within-configuration gaps for the full subsets.
Feature availability differs across configurations, so the stable-run sets
are not identical. Table~\ref{tab:mismatch_k5_overlap} reports the results.
We observed that curvature features add little discrimination (within-program \ROCAUC 0.816 to
0.816, cross-program 0.730 to 0.731).
Optimizer features substantially improve within-program detection (0.899 to
0.971) but slightly reduce cross-program performance relative to the baseline model (0.826 to 0.782). The gap between within and held for the $+$\OF condition is 0.189.

% --- Table: Mismatch overlap-controlled ---
\begin{table}[htbp]
  \centering\footnotesize
  \caption{Overlap-controlled \tvmm detection (\kb{5})}
  \label{tab:mismatch_k5_overlap}
  \setlength{\tabcolsep}{2.5pt}
  \resizebox{\columnwidth}{!}{%
  \begin{tabular}{@{}llcrrll@{}}
    \toprule
    \headrow
    \textbf{Config} & \textbf{Features} & \textbf{Setting}
      & $n$ & $\pi$
      & \textbf{\ROCAUC}
      & \textbf{\PRAUC} \\
    \midrule
    \multirow{4}{*}{\CF}
      & Baseline  & within  & 1317 & .50
        & 0.816 \ci{0.618}{0.981}
        & 0.855 \ci{0.659}{0.973} \\
      & Baseline  & cross   & 1317 & .50
        & 0.730 \ci{0.475}{0.959}
        & 0.773 \ci{0.441}{0.921} \\
      & $+$\CF    & within  & 1317 & .50
        & 0.816 \ci{0.618}{0.981}
        & 0.856 \ci{0.661}{0.974} \\
      & $+$\CF    & cross   & 1317 & .50
        & 0.731 \ci{0.476}{0.959}
        & 0.774 \ci{0.442}{0.921} \\
    \midrule
    \multirow{4}{*}{\OF}
      & Baseline  & within  & 410 & .42
        & 0.899 \ci{0.704}{0.987}
        & 0.907 \ci{0.637}{0.988} \\
      & Baseline  & cross   & 410 & .42
        & 0.826 \ci{0.596}{0.970}
        & 0.830 \ci{0.491}{0.978} \\
      & $+$\OF   & within  & 410 & .42
        & 0.971 \ci{0.860}{0.999}
        & 0.957 \ci{0.817}{1.000} \\
      & $+$\OF   & cross   & 410 & .42
        & 0.782 \ci{0.434}{0.988}
        & 0.803 \ci{0.444}{0.987} \\
    \bottomrule
  \end{tabular}%
  }\\[2pt]
  \parbox{\columnwidth}{\scriptsize\raggedright
  $n$ = stable runs, $\pi$ = positive-class prevalence.}
\end{table}

\textit{Full-dataset analysis.}
Table~\ref{tab:mismatch_k5} reports the same analysis on the full
program subsets. The \CF\ subset has 3,278 stable runs across 22 programs at prevalence
0.37, and the \OF\ subset has 1,256 stable runs across 26 programs at prevalence 0.51.
The two subsets cover different programs, so we focus on
within-configuration generalizability gaps.
Curvature features add negligible discrimination (within-program 0.799 to 0.799, cross-program
0.738 to 0.738).
Optimizer features improve within-program detection (0.874 to 0.897) but
degrade substantially under program-held-out evaluation (0.858 to 0.612), a
generalizability gap of 0.285.
The larger gap in the full dataset relative to the overlap-controlled
subset (0.285 vs.\ 0.189) reflects sensitivity to program composition, as the two subsets cover different programs with different mismatch prevalences (Table~\ref{tab:mismatch_k5}).

% --- Table: Mismatch full dataset ---
\begin{table}[htbp]
  \centering\footnotesize
  \caption{\tvmm detection at \kb{5} on full dataset (stable runs only)}
  \label{tab:mismatch_k5}
  \setlength{\tabcolsep}{2.5pt}
  \resizebox{\columnwidth}{!}{%
  \begin{tabular}{@{}llcrrll@{}}
    \toprule
    \headrow
    \textbf{Config} & \textbf{Features} & \textbf{Setting}
      & $n$ & $\pi$
      & \textbf{\ROCAUC}
      & \textbf{\PRAUC} \\
    \midrule
    \multirow{4}{*}{\CF}
      & Baseline     & within  & 3278 & .37
        & 0.799 \ci{0.657}{0.918}
        & 0.794 \ci{0.611}{0.914} \\
      & Baseline     & cross   & 3278 & .37
        & 0.738 \ci{0.577}{0.871}
        & 0.632 \ci{0.410}{0.825} \\
      & $+$\CF       & within  & 3278 & .37
        & 0.799 \ci{0.657}{0.919}
        & 0.794 \ci{0.611}{0.914} \\
      & $+$\CF       & cross   & 3278 & .37
        & 0.738 \ci{0.577}{0.872}
        & 0.632 \ci{0.411}{0.825} \\
    \midrule
    \multirow{4}{*}{\OF}
      & Baseline     & within  & 1256 & .51
        & 0.874 \ci{0.681}{0.972}
        & 0.922 \ci{0.709}{0.987} \\
      & Baseline     & cross   & 1256 & .51
        & 0.858 \ci{0.671}{0.949}
        & 0.892 \ci{0.667}{0.974} \\
      & $+$\OF      & within  & 1256 & .51
        & 0.897 \ci{0.728}{0.979}
        & 0.916 \ci{0.703}{0.987} \\
      & $+$\OF      & cross   & 1256 & .51
        & 0.612 \ci{0.371}{0.846}
        & 0.566 \ci{0.267}{0.873} \\
    \bottomrule
  \end{tabular}%
  }\\[2pt]
  \parbox{\columnwidth}{\scriptsize\raggedright
  $n$ = stable runs, $\pi$ = positive-class prevalence. Optimizer features show a cross-program generalizability gap of 0.285.}
\end{table}

\begin{figure}[htbp]
  \centering
  \subfloat[Curvature features (\CF)\label{fig:mismatch_sweep_cf}]{%
    \includegraphics[width=0.50\linewidth]{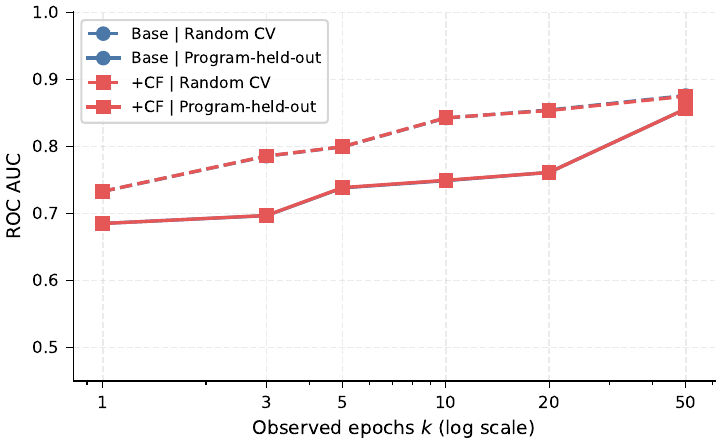}%
  }\hfil
  \subfloat[Optimizer features (\OF)\label{fig:mismatch_sweep_of}]{%
    \includegraphics[width=0.50\linewidth]{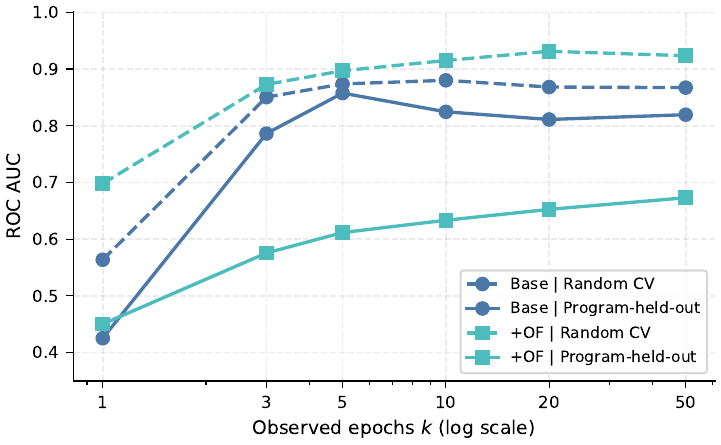}%
  }
  \caption{\tvmm ROC-AUC across observation windows}
  \label{fig:mismatch_sweep}
\end{figure}

\begin{figure}[htbp]
  \centering
  \includegraphics[width=0.90\linewidth]{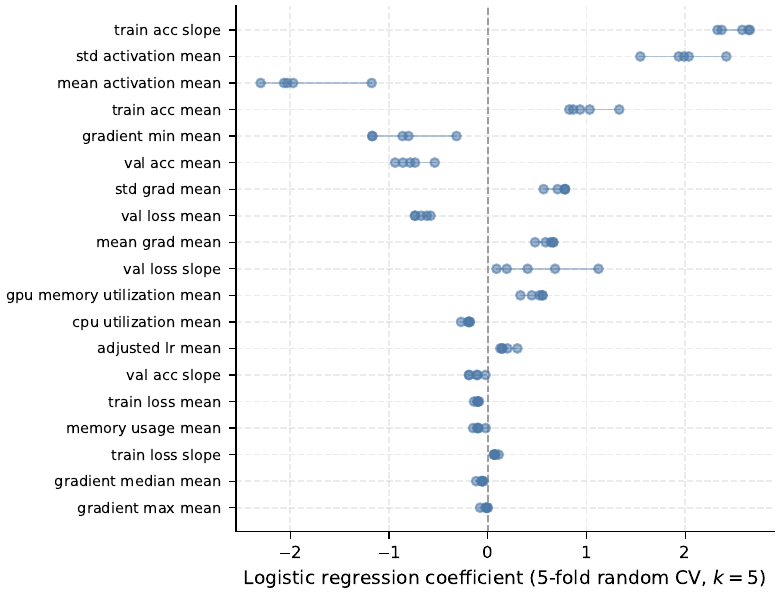}
  \caption{Optimizer-feature logistic regression coefficients across within-program CV folds}
  \label{fig:csi_coeff}
\end{figure}

Fig.~\ref{fig:csi_coeff} shows that the linear model's optimizer-feature
coefficients vary substantially across within-program CV folds,
which suggests that these features are being used through program-specific
patterns rather than shared fault semantics. This pattern appears consistent with shortcut learning~\cite{geirhos2020shortcut}, where classifiers rely on correlations that hold in-distribution but not under shift. This interpretation is consistent with our cross-program results, where activation and optimizer features degraded on unseen programs with different characteristic scales.

\begin{figure}[htbp]
  \centering
  \includegraphics[width=0.90\columnwidth]{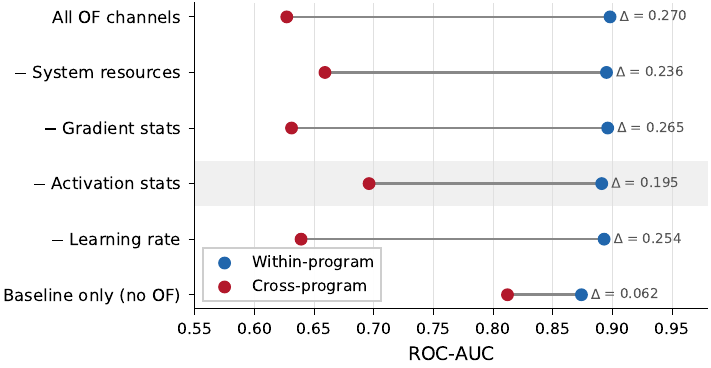}
  \caption{Feature-group ablation for \tvmm detection at \kb{5}}
  \label{fig:ablation}
\end{figure}

\textit{Feature-group ablation.}
To identify which optimizer-feature groups are most associated with the
cross-program degradation, Fig.~\ref{fig:ablation} reports \tvmm \ROCAUC at \kb{5}
under program-held-out evaluation after removing each feature group
in turn from the full \OF channel set (1,256 stable runs,
26~programs).
We found that removing activation statistics leads to the largest cross-program
improvement (0.627 to 0.696, gap reduction from 0.270 to 0.195),
suggesting that per-layer activation scales are a primary source
of program-specific shortcut patterns.
Removing system resource metrics also narrows the gap (0.270 to 0.236).
In contrast, removing gradient statistics or learning rate has
minimal effect on cross-program performance.
The baseline-only condition (i.e., no \OF channels) achieves the
smallest gap (0.062) but retains cross-program \ROCAUC of 0.812,
suggesting that loss and accuracy metrics alone may provide the most
reliable mismatch detection across programs.

% --- Figure: Feature-group ablation ---

\begin{rqbox}
\textbf{RQ$\mathbf{_3}$ Summary.}
We found that optimizer features substantially improve within-program \tvmm detection but degrade under program-held-out evaluation (generalizability gap up to 0.285), driven primarily by per-layer activation statistics. These results reveal an observability--generalizability trade-off introduced by richer runtime features that is not visible under within-program evaluation.
\end{rqbox}

\subsection{Re-evaluation of Existing Techniques}
\label{sec:comparison}

To assess whether the evaluation-strategy gap generalizes beyond our
linear methodology, we reimplement the feature extraction code
of four existing dynamic-analysis
techniques (DEFault~\cite{jahan2025default},
DeepFD~\cite{cao2022deepfd},
AutoTrainer~\cite{autotrainer}, and
DeepDiagnosis~\cite{wardat2022deepdiagnosis}) and evaluate them on
\dataset under both settings on the four-class fault-type task at \kb{5}.
We follow the feature definitions in the original papers~\cite{cao2022deepfd,autotrainer,jahan2025default,wardat2022deepdiagnosis} and use their provided replication packages. We reproduced the reported accuracy values to within 3\%~\cite{icsmeReplicationPackage}.

\begin{figure}[htbp]
  \centering
  \includegraphics[width=\linewidth]{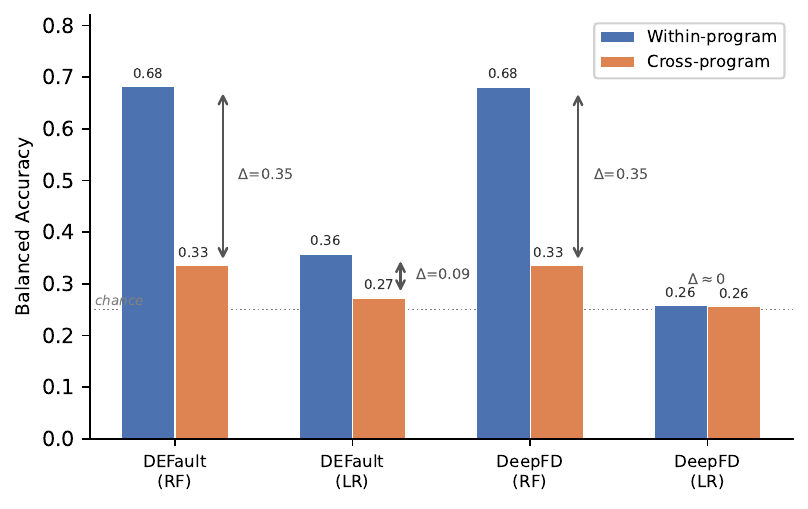}
  \caption{Within- vs.\ cross-program accuracy for DEFault and DeepFD at \kb{5}}
  \label{fig:comparison_ba_gap}
\end{figure}

\textbf{Learning-based techniques.}
DEFault features (42 features) and DeepFD features (160 features) are
evaluated with both Random Forest (RF) and Logistic Regression (LR).
Fig.~\ref{fig:comparison_ba_gap} reports the results. With their original RF classifiers, both techniques achieve
within-program \BA above 0.65 but cross-program \BA drops to
0.32, a gap of 0.34. This gap is \emph{larger} than the 0.190 observed with
logistic regression in RQ1, which suggests that the more expressive RF
classifier captures more program-specific structure.
Under LR, DEFault features show a gap of 0.088 (within 0.357, cross 0.270), while DeepFD features
show a near-zero gap at near-chance
performance, which suggests that DeepFD's 160 statistical features are
largely non-discriminative under a linear model.

\textbf{Rule-based techniques.}
AutoTrainer and DeepDiagnosis are rule-based detectors that identify the
\emph{presence} of training problems (e.g., vanishing gradient, dying ReLU)
but do not classify the underlying fault type.
AutoTrainer's five rules trigger on 94.2\% and
DeepDiagnosis's eight symptom detectors on 100\% of traces
in \dataset, indicating that the traces contain the monitored runtime metrics.
The rules fire on all fault types without distinction, so they cannot
produce fault-type classification metrics, and
the evaluation-strategy distinction does not apply to fixed rules.

We found that the evaluation-strategy gap is not an artifact of our linear
methodology. It is larger, not smaller, with the higher-capacity RF
classifiers used by existing techniques.
In our experiments, the gap therefore appears to reflect the cross-program
generalization challenge itself rather than a specific classifier or feature
set.

% -----------------------------------------------------------------------

\subsection{Discussion and Implications}

Our results indicate that evaluation strategy and runtime feature design interact, jointly influencing both
measured and actual diagnostic performance.
Under within-program CV, we observed that all feature sets benefit from program-specific
structure, so their apparent performance levels converge.
Only under program-held-out evaluation did the generalizability
differences between feature designs become visible in our experiments. This interaction has direct implications for feature design.
Optimizer features improve within-program mismatch detection
(\ROCAUC 0.90 to 0.97 on the overlap-controlled subset), but this
improvement degrades under program-held-out evaluation (0.78),
exposing program-specific shortcuts.
Curvature features, by contrast, provided a generalizable instability signal
under both strategies in our experiments. We observed that features that appear similarly effective within a program pool
may differ substantially when program-specific structure cannot be exploited. The appropriate instrumentation therefore depends on the task. Loss and accuracy features are often sufficient for mismatch detection, whereas curvature features are worth their additional cost when early instability must not be missed. We derive the following implications from these findings.
\begin{enumerate}[leftmargin=*,label=\arabic*.,nosep]
  \item \textbf{Align validation with deployment.} Cross-program claims require program-held-out evaluation~\cite{zimmermann2009cpdp,lyu_tosem2021}, since within-program results may overestimate cross-program performance. Reporting program-identity predictability serves as an indicator of program-specific shortcuts.
  \item \textbf{Evaluate feature configurations separately.} Additional logging channels can introduce shortcuts detectable only under grouped evaluation~\cite{sculley2015debt}. Each feature configuration should be evaluated under program-held-out splits before deployment. Our results indicate that per-program normalization reduces program-identity predictability without degrading cross-program performance, suggesting that normalization-invariant designs may improve transfer.
  \item \textbf{Monitor instability at initialization.} In our dataset, 96\% of instability occurred at epoch~0, and curvature features generalized under program-held-out evaluation, supporting a high-precision early-termination rule, while supervised models offer adjustable precision--recall in this setting.
\end{enumerate}

% -----------------------------------------------------------------------
\section{Threats to Validity}
\label{sec:threats}
\textit{External validity.}
Our findings derive from 38 programs in \dataset with
mutation-injected faults.
Mutation enables controlled experimentation and the fault
categories are grounded in empirical bug taxonomies
\cite{humbatova2020taxonomy,deepcrime}.
Mutants may not capture the full diversity of real-world
training failures. We mitigate optimism by adopting program-held-out evaluation, which explicitly tests generalization to unseen programs, and we treat replication on real-fault corpora as future work to confirm that the gap magnitudes hold beyond mutation-injected faults.

\textit{Construct validity.}
Fault-type labels correspond to the applied mutation operators.
The \tvmm label relies on the corpus-provided
\texttt{acc\_gap\_too\_big} flag with a fixed threshold~\cite{jahan2025default}.
We fixed this threshold to match the DEFault corpus construction protocol and did not tune it per architecture.
RNNs, CNNs, and FFNNs differ in expected training/validation gap dynamics, so threshold sensitivity across architecture types should be evaluated in future work.
Alternative operationalization may produce different absolute metrics,
but the central observation that
within-program evaluation reflects program-specific feature structure
does not depend on the particular threshold.

\textit{Internal and conclusion validity.}
We use linear models to reduce the risk that classifier capacity
dominates the observed effects.
Confidence intervals are computed via a cluster bootstrap over programs~\cite{cameron2008bootstrap}.
With 38 clusters, uncertainty is non-trivial, so we report full
intervals and avoid over-interpreting small differences.
The \CF and \OF configurations were collected on partially different program subsets. To reduce this confound, we report both overlap-controlled comparisons on shared programs and within-configuration generalizability gaps on the full subsets. Absolute \CF-vs-\OF performance levels should therefore be interpreted cautiously.

\section{Related Work}
\label{sec:related}

Empirical studies of DL bugs characterize common root causes and failure symptoms, and provide the taxonomies that diagnosis and mutation operators build on~\cite{islam2019bugs,humbatova2020taxonomy,yang2022dlframeworkbugs,zhang2018tensorflowbugs}.
Building on these taxonomies, dynamic-analysis techniques learn from runtime training metrics to detect and diagnose failures.
DeepDiagnosis~\cite{wardat2022deepdiagnosis} and AutoTrainer~\cite{autotrainer} implement symptom detectors and repair heuristics for training problems.
DeepFD~\cite{cao2022deepfd} and DEFault~\cite{jahan2025default} frame diagnosis as supervised learning over aggregated runtime metrics, allowing fault-type classification and localization. Qi et al.~\cite{qi2024coverage} further augment runtime metrics with coverage-derived metrics.
Across this line of work, evaluations measure within-program reuse and, in several cases, detection on real-world fault benchmarks. The cross-program generalization challenge, where the diagnostic model encounters a program with different architecture, dataset, and training dynamics, has not been separately measured. Additional techniques support interactive diagnosis and pre-execution checks. Cockpit~\cite{schneider2021cockpit} and UMLAUT~\cite{schoop2021umlaut} expose gradient distributions, curvature-related signals, and program structure to support human-in-the-loop debugging.
DeepLocalize~\cite{wardat2021deeplocalize} uses dynamic analysis of value propagation between layers to localize faults in DNNs.
Hessian-derived metrics have also been used to localize fault sources in attention models~\cite{jahan2025hessian}.
Static analyses for numerical bugs~\cite{zhang2020debar} and design-by-contract approaches for DL APIs~\cite{ahmed2023dlcontract} aim to prevent certain failures before execution.

Many diagnosis studies rely on mutation-injected faults to obtain labeled data at scale.
DeepMutation~\cite{deepmutation} and DeepCrime~\cite{deepcrime} design mutation operators grounded in empirical bug taxonomies.
A recent study shows that pre-training mutants better match real faults than post-training mutants under coupling and behavioral-similarity criteria~\cite{ahmed2025mutantrealism}, supporting the construct validity of fault-injected corpora.
\dataset follows this approach and uses mutation-injected runs to enable controlled measurement of generalization across programs. The need to align validation with deployment is well established in cross-project defect prediction~\cite{zimmermann2009cpdp,herbold2018comparative,tantithamthavorn2017validation}, grouped cross-validation for clustered data~\cite{roberts_ecography2017}, and operational ML splitting decisions~\cite{lyu_tosem2021,sculley2015debt,amershi2019seml}.
Separately, shortcut learning~\cite{geirhos2020shortcut} explains why rich runtime features can degrade under distribution shift, and curvature-aware analyses~\cite{martens2010,cohen2021edge,arora2022edge,gilmer2022curvature,pearlmutter1994} motivate curvature-derived signals as transferable instability indicators.
Our study connects these threads by measuring the evaluation-strategy gap for DL fault diagnosis and showing that evaluation protocol and feature design interact to determine both measured and actual diagnostic performance.

\section{Conclusion}
\label{sec:conclusion}

In this work, we studied whether dynamic-analysis fault diagnosis techniques for DL training failures generalize to programs that were not seen during training. Using \dataset, we found that fault-type diagnosis achieved higher balanced accuracy under within-program evaluation than under program-held-out evaluation, with a gap of 0.190. We also observed a similar performance gap across four existing fault diagnosis techniques. Further analysis indicates that program-level structure in the runtime features explains much of this gap, while some fault-related information still generalizes across programs. Our feature analysis shows that gains from richer runtime features do not always transfer to unseen programs. Curvature features helped detect instability on unseen programs. Optimizer and activation features improved training/validation mismatch detection mainly within the same program, with much smaller gains on unseen programs. These findings suggest that within-program evaluation can overestimate diagnostic performance and make richer logging appear more useful than it is for cross-program diagnosis. Future work should assess whether these patterns hold on real-world faults and design methods that help diagnostic models focus on fault-related behavior rather than program-specific runtime patterns.
% -----------------------------------------------------------------------
\printbibliography

\end{document}